\documentclass[epj-spec]{svjour}
\usepackage{graphics}
\providecommand{\openone}{\leavevmode\hbox{\small1\kern-3.8pt\normalsize1}}
\begin{document}
\title{Memory effects in quantum information 
        transmission across a Hamiltonian dephasing channel}
\author{A. D'Arrigo \inst{1}\fnmsep\thanks{\email{antonio.darrigo@dmfci.unict.it}} 
   \and G. Benenti \inst{2,3}\fnmsep\thanks{\email{giuliano.benenti@uninsubria.it}}
   \and G. Falci  \inst{1}\fnmsep\thanks{\email{gfalci@dmfci.unict.it}} }
\institute{MATIS CNR-INFM, Catania \& Dipartimento di Metodologie Fisiche e 
            Chimiche per l'Ingegneria, Universit\`a degli Studi di Catania,  
            Viale Andrea Doria 6, 95125 Catania, Italy 
     \and CNISM, CNR-INFM \& Center for Nonlinear and Complex Systems, 
            Universita' degli Studi dell'Insubria, Via Valleggio 11, 22100 Como, Italy
     \and Istituto Nazionale di Fisica Nucleare, Sezione di Milano,
            via Celoria 16, 20133 Milano, Italy}
\abstract{We study a dephasing channel with memory, modelled by a 
multimode environment of oscillators. Focusing on the case of two channel
uses, we show that memory effects can enhance the amount of 
coherent quantum information transmitted down the channel.
We also show the Kraus representation for two channel uses.
Finally, we propose a coding-decoding scheme that takes 
advantage of memory to improve the fidelity of transmission.}

\maketitle


\section{Introduction}

Quantum communication 
channels \cite{kn:nielsen-chuang,kn:benenti-casati-strini} 
use quantum systems to transfer
classical or quantum information. In the first case, 
classical bits are encoded by means of quantum states. 
In the latter, one
can transfer an unknown quantum state between different units
of a quantum computer, or  
distribute entanglement between two or more communicating parties. 
The fundamental quantities characterizing a quantum channel 
are the \textit{classical} and the \textit{quantum channel capacities},
that are defined as the maximum number of bits/qubits that can be reliably 
transmitted per channel use \cite{BennettShor98}. 

Quantum channels are the natural theoretical framework to investigate
both quantum communication and computation in a noisy environment.
In the first case, information is transmitted in space, in the latter
in time. In both cases, noise can have relevant low frequency components,
which traduce themselves in memory effects. That is, consecutive uses of
a channel can be correlated.
Memory effects may be important, for instance, in 
quantum communication protocols realized by means of photons 
travelling across fibers with 
birefringence fluctuating with characteristic time scales longer than the 
separation between successive light pulses \cite{Banaszek04}. 
Moreover, solid state 
implementations of quantum hardware show a characteristic
low-frequency noise \cite{kn:paladino-1}. 

Previous investigations have shown that memory can enhance 
classical information transmission down a quantum memory channel 
\cite{MacchiavelloPalma02}.
More recently, we have considered 
a channel subject to dephasing noise described by a 
Markov chain with memory effects, showing 
that the quantum capacity  
increases with respect to the the memoryless 
case \cite{DarrigoBenentiFalci07}.
Furthermore, based on theoretical arguments and 
numerical simulations, we have conjectured that the 
enhancement of the quantum capacity also
takes place for a dephasing  quantum  environment 
modelled by a bosonic bath \cite{DarrigoBenentiFalci07}.
In this paper, we discuss memory effect in 
this latter model and address the case of two channel uses.
This is interesting because, while channel capacity is 
defined in the asymptotic limit of an infinite number of 
uses, real coding-decoding 
schemes \cite{kn:nielsen-chuang,kn:benenti-casati-strini} 
necessarily work on a finite number of uses.
The case of two channel uses nicely show that 
memory effects may enhance the coherent quantum information 
transmitted down the channel. Furthermore, we propose a 
simple coding-decoding scheme that takes 
advantage of memory to improve the fidelity of transmission.

Dephasing channels are characterized by the property that when N qubits 
are sent through the channel, the states of a preferential orthonormal basis 
$\{|j\rangle \equiv |j_1,....,j_N\rangle, \,j_1,...,j_N=0,1\}$
are transmitted without errors. 
Therefore, dephasing channels 
are noiseless from the viewpoint of the 
transmission of classical information, since the states of the
preferential basis can be used for encoding classical information. 
Of course superpositions of basis states may decohere, 
thus corrupting the transmission of quantum information. 
We point out that dephasing channels are relevant for systems 
in which relaxation is much slower than dephasing 
\cite{Leung99,kn:inhom}. 

The paper is organized as follows.
In Sec.~2 we review basic quantities useful to describe quantum 
information transmission across a noisy channel.
In Sec.~3 we introduce the Hamiltonian dephasing channel
investigated in the paper, while in Sec.~4 we recall 
some known results for a single channel use, relevant 
to describe the memoryless limit. 
Finally, in Sec.~5 we study the case of two 
channel uses. In particular, we report the channel transformation
for a generic input (Sec.~\ref{sec-5.1}) and the 
corresponding Kraus representation (Sec.~\ref{sec-5.2}).
We also discuss memory effects for a class of input 
states (Sec.~\ref{sec-5.3}) and finally propose a simple and 
efficient coding/decoding scheme that takes advantage of 
memory (Sec.~\ref{sec-5.4}). Our conclusions are in Sec.~6.

\section{Quantum information transmission: basic definitions}
\label{sec-1}
In this section, we briefly review basic quantities and concepts
that are useful to describe the channel ability to transmit quantum information.
For this purpose, we consider as quantum information carrier a quantum system
$\cal Q$ and describe the channel action on $\cal Q$ by means of a 
superoperator \cite{kn:nielsen-chuang,kn:benenti-casati-strini} $\cal E$,
that is, a completely positive, trace preserving 
linear map that transforms the (generally mixed) input state 
$\rho^{\cal Q}$ into the output state $\rho^{\cal Q'}$.
In a prototype communication protocol, ${\cal Q}$ may be an $N$-qubit
system representing $N$ successive uses of the channel.

\subsection{Quantum information}
\label{sec-2}
Quantum information is the information related to a quantum
source \cite{Schumacher95,BarnumNielsenSchumacher98}, 
that is, a source $\Sigma$ of identical quantum systems $\cal Q$, which are
prepared in an unknown quantum state chosen from the ensemble $\{\rho_k\}$,
according to a given stationary probability distribution $\{p_k\}$.
The amount of information
generated by the source is measured by the von Neumann entropy 
$S(\Sigma) \equiv S(\rho)=
-\mathrm{Tr}\big[\rho\log_2\rho\big]$, where
$\rho=\sum_k p_k \rho_k$
\cite{Schumacher95,BarnumNielsenSchumacher98}. Indeed, $S(\Sigma)$ 
provides the number of quantum two level systems necessary  
to efficiently encode the source $\Sigma$ according 
to the noiseless quantum coding theorem.  
Quantum information reduces to classical information only if the 
states $\{\rho_k\}$ are pure and mutually orthogonal.

\subsection{Reliable transmission and entanglement fidelity}
\begin{figure} 
 \resizebox{1\columnwidth}{!}{\includegraphics{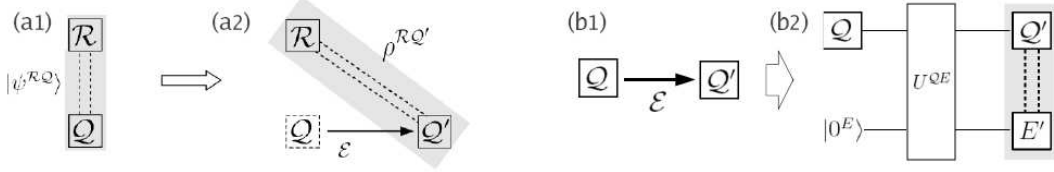}}
 \caption{Left: The system ${\cal Q}$ is considered as 
              entangled with a reference system ${\cal R}$, so that 
              the initial state $|\psi^{\cal RQ }\rangle$ of
              the overall system ${\cal RQ }$ is pure;
              then (a2) the system ${\cal Q}$ is sent through 
              the channel which affects both the system ${\cal Q}$
              and the entanglement between ${\cal Q}$ and ${\cal R}$.
 	  Right: Sketch of the equivalence between the map $\cal E$ (b1) and
              the unitary representation in (b2), in which the effect 
              of the map $\cal E$ on the system is reproduced by a 
              unitary evolution $U^{{\cal Q}E}$ of the system itself 
              plus a fictitious environment $E$, which is 
              initially in a pure state.
          }
  \label{Ef-Se}
 \end{figure}

A proper way to measure reliability of quantum information
transmission is the entanglement fidelity \cite{Schumacher96,BarnumKnillNielsen00}.
To define this quantity we look at the system ${\cal Q}$ 
as a part of a larger quantum system ${\cal RQ }$, 
initially in a pure entangled  state $|\psi^{\cal RQ } \rangle$.  
The density operator of the system ${\cal Q }$ 
is then obtained from that of ${\cal  RQ}$  by a partial trace over ${\cal R}$:
$\rho^{\cal Q}=\mathrm{Tr}_{\cal R} [|\psi^{\cal RQ } 
\rangle  \langle \psi^{\cal RQ } | ]$.
The system ${\cal Q}$ is sent through the channel and undergoes
the transformation ${\cal E}$, while ${\cal R}$ is ideally isolated 
from the environment (see Fig. (\ref{Ef-Se}) left).
The final state of the composite system is:
 \begin{equation}
 	 \rho^{{\cal RQ}'}= ({\cal I}^{\cal R} 
	 \otimes {\cal E}^{\cal Q})\Big(  |\psi^{\cal RQ }
	 \rangle \langle \psi^{\cal RQ } | \Big),
  \label{rhoRQ'}
 \end{equation}
where ${\cal I}$ is the identity superoperator.
 Let us consider the following question: \textit{how faithfully 
 does the channel preserve
 the entanglement} between the two systems $\cal Q$ and $\cal R$? 
 The answer is the \textit{entanglement fidelity} $F_e$
 \cite{Schumacher96}, defined as
 the fidelity $F$ between the  initial pure state $|\psi^{\cal RQ } \rangle$ 
 and the final (generally mixed) state $\rho^{{\cal RQ}'}$:
 \begin{eqnarray}
	F_e\,= \, F_e(\rho^{\cal Q},{\cal E})\,=
        \,F\big(|\psi^{\cal RQ } \rangle,\rho^{{\cal RQ}'}\big)\,=\,
        \langle \psi^{\cal RQ } |\,
        ({\cal I}^{\cal R} \otimes {\cal E}^{\cal Q})
        \big(|\psi^{\cal RQ } \rangle \langle \psi^{\cal RQ }| \big)
        \,|\psi^{\cal RQ } \rangle.
  \label{Fe_def_1}
 \end{eqnarray}
 It can be shown that entanglement fidelity only depends 
 on the initial state $\rho ^{\cal Q}$ of the system and on the channel
 action ${\cal E}$, not on the particular purification 
 $|\psi^{\cal RQ } \rangle$ chosen \cite{Schumacher96}.
 Unlike the usual input-output fidelity 
 \cite{kn:nielsen-chuang,kn:benenti-casati-strini},
 entanglement fidelity looks at the same physical process,
 the transmission of $\cal Q$ across a channel, from a different point
 of view: as a local transformation on a part ($\cal Q$) of an entangled
 system ($\cal RQ$). This transformation is undesired 
 because it can reduce the amount of entanglement of the overall system.
 
 \subsection{Channel noise and entropy exchange}
 The entropy exchange \cite{Schumacher96} 
 is the entropy that the enlarged system
 ${\cal RQ}$ 
 acquires when {$\cal Q$} undergoes the transformation ${\cal E}$:
 \begin{equation} 
    S_e\,=S_e(\rho^{\cal Q},{\cal E})\,=\,S(\rho^{{\cal RQ}'}),
  \label{entropy exchange}
 \end{equation}
 where $\rho^{{\cal RQ}'}$ is given by (\ref{rhoRQ'}).
 It can be shown that the entropy exchange 
 is an intrinsic function of the system input $\rho^{\cal Q}$ 
 and of the channel $\cal E$ and does not depend on 
 the particular purification\cite{Schumacher96}.
 Since $\cal E$ is a completely positive trace preserving linear map, 
 it can be represented \cite{kn:nielsen-chuang,kn:benenti-casati-strini} 
 by a unitary evolution $U^{{\cal Q}E}$ on a larger system, given 
 by the system $\cal Q$ itself plus an ancilla $E$,
 that is, a (generally fictitious) environment initially in a pure state
 (see Fig. (\ref{Ef-Se}) right). 
 It is straightforward to show that \cite{Schumacher96,SchumacherNielsen96}:
 \begin{equation}   
   S(\rho^{{\cal RQ}'})=S(\rho^{E'})=S_e,
 \end{equation}
 where $\rho^{E'}$ is the final state of the environment. 
 $S_e$ measures the entropy increase of the environment $E$ or, equivalently, 
 the entanglement between $\cal RQ$ and $E$ after the evolution $U$.
 The entropy exchange 
 can be thought of as the quantum analogue of the classical 
 \textit{conditional entropy} \cite{Shannon48,CoverThomas91}, 
 that measures the average 
 uncertainty about the channel output for a known input,
 namely the noise added to the output by the channel. 

 \subsection{Quantum capacity and coherent information}
Classical channel capacity is given by 
the maximum (over the input probability distribution) of 
the input-output \textit{mutual information} 
\cite{Shannon48,CoverThomas91}.
The quantity analogous to mutual information for quantum information is  
the \textit{coherent information} $I_c$
\cite{SchumacherNielsen96,Lloyd97}, defined as
 \begin{equation}
	I_c\,=\,I_c\big(\rho^{\cal Q},{\cal E}\big) \,=\,
 	S\big({\cal E}(\rho^{\cal Q})\big) \, - \,
        S_e( \rho^{\cal Q},{\cal E}) \,=\, 
        S\big(\rho^{{\cal Q}'}\big) \, - \, S( \rho^{{\cal RQ}'}).
	\label{CoherentInformation}
 \end{equation} 
For memoryless channels and for the so-called forgetful channels
\cite{KretschmannWerner05},
for which memory effects decay exponentially with time, the
\textit{quantum channel capacity}  
$Q$ is the maximum of $I_c/N$ over all possible input states,
in the limit of number of channel uses $N\to\infty$
\cite{barnum,shor,devetak,winter}.
Note that the coherent information is maximal when ${\cal Q}$ and 
${\cal R}$ are maximally entangled and the channel is noiseless. 
Indeed, in this case $S\big(\rho^{\cal Q'}\big)=S\big(\rho^{\cal Q}\big)$
is maximal because the input state $\rho^{\cal Q}$ is maximally mixed, 
and $S\big( \rho^{{\cal RQ}'}\big)=0$, since the state 
$|\psi^{\cal RQ}\rangle$ remains pure after the transmission.  
Smaller values of the coherent information are obtained when 
the state $|\psi^{\cal RQ}\rangle$ is not maximally entangled
or noise affects the channel.  
This example illustrates the fact that coherent information measures
the possibility to convey entanglement through a communication channel.

\section{Channel Hamiltonian model}
\label{sec-3}
We suppose that information is carried by qubits that transit across 
a communication channel, modeled as a purely dephasing 
environment. 
The Hamiltonian describing the transmission of $N$ qubits through 
the channel reads \cite{DarrigoBenentiFalci07}
 \begin{equation}
   H(t) = H_{E} -\frac{1}{2}{X}_{E}{F}(t), \quad
   F(t) =\lambda
              \sum_{k=1}^{N}\sigma_z^{(k)}\, f_k(t),
  \label{Hamiltonian}
 \end{equation}
where $ H_{E}$ is the environment Hamiltonian and 
${X}_{E}$ the environment coupling operator. 
The $k-$th qubit is coupled to the environment via 
its Pauli operator $\sigma_z^{(k)}$, the coupling strength
being $\lambda$.
The functions $f_k(t)$ switch on and off the coupling:  
$f_k(t)=1$ when the $k$-th qubit is inside the channel, 
$f_k(t)=0$ otherwise. 
We call $\tau_p$ the time each
carrier takes to cross the channel and $\tau$ the
time interval that separates
two consecutive qubits entering the channel.
Note that Hamiltonian (\ref{Hamiltonian}) is expressed in 
the interaction picture with respect to the qubits. 

We call $\omega_0$ and $\rho^{\cal Q}$ the density operators which 
represent the initial states of the environment and of the 
$N$ qubits, respectively. 
Assuming that initially the system and the environment are 
not entangled, we can write the state of the system at time $t$ as 
follows: 
\begin{equation} 
\rho^{\cal Q}(t)=
\mathrm{Tr}_{E}\{U(t) (\rho^{\cal Q} \otimes \omega_0) U^{\dag}(t)\},
 \label{system_final_state_01} 
\end{equation}
where
 \begin{equation} 
 U(t) \,=\, {T} e^{-\frac{i}{\hbar} \int_{0}^t ds \,H(s)}.
  \label{evolution_operator_01}
 \end{equation}
In particular, we are interested in the final state $\rho^{\cal Q'}$ 
after all $N$ qubits crossed the channels.
To treat this problem we choose the factorized basis states
$\{|j \, \alpha_{E}\rangle\}$, where 
$\{|j \rangle = |j_1,...,j_N\rangle\}$ are the
eigenvectors of $\prod_k \sigma_z^{(k)}$, and 
$\{|\alpha_E \rangle\}$ is an orthonormal basis for the environment.
The dynamics preserves the basis states $|j\rangle$ and therefore
the evolution operator (\ref{evolution_operator_01}) is
diagonal in the system indices:
\begin{equation} 
\langle j \,\alpha_{E} | U(t) | l \alpha_{E}^\prime \rangle 
= \delta_{jl} \,
\langle \alpha_{E}| U(t|j) |\alpha^\prime_{E} \rangle,
\label{evolution_operator_02}
\end{equation}
where $U(t|j)=\langle j | U(t) | j\rangle$ expresses the
conditional evolution operator of the environment alone.
Therefore, 
\begin{equation}
(\rho^{\cal Q'})_{jl}= 
(\rho^{\cal Q})_{jl}\sum_\alpha \langle \alpha_{E} | \, U(t|j) \,w_0\,
U^\dagger (t|l) \,|\alpha_{E}\rangle.
\label{eq:spinbosonrho}
\end{equation}
In this basis representation the populations are preserved and
the environment only changes the off-diagonal elements of $\rho^{\cal Q}$.

Now we model the environment with an infinite set of 
oscillators:
 \begin{equation} 
    {H}_{E}=\sum_\alpha \omega_\alpha {b}^\dag_\alpha {b}_\alpha \,+\, H_C,
\quad
H_C=\sum_\alpha \frac{\lambda^2}{4\omega_\alpha}
\sum_{k=1}^N \sigma_z^{(k)}, 
\quad
X_E=\sum_\alpha(b^\dag_\alpha+b_\alpha),
 \label{multimode_environment}
 \end{equation}
where $H_C$ is a counterterm \cite{weiss}.
If the environment is initially in thermal equilibrium, 
$w_0=e^{-\beta {\cal H}_{E}}$,
we obtain \cite{DarrigoBenentiFalci07}
\begin{equation}
\sum_\alpha \langle \alpha_{E} | \, U(t|j) 
\,w_0 \,U^{\dagger}(t|l)\,|\alpha_{E} \rangle  
=
      \exp\Big[-{\lambda^2} \hskip-5pt \int\limits_0^{\;\;\infty}
      \hskip-3pt\frac{d\omega}{\pi}
      S(\omega)\frac{1-\cos(\omega \tau_p)}{\omega^2}
      \Big| \sum_{k=1}^N(j_k-l_k) 
      e^{i\omega(k-1)\tau}\Big|^2\Big],
 \label{general_dephasing_factor}
\end{equation}
where $S(\omega)$ is the power spectrum of the coupling 
operator $X_{E}$, that is the Fourier transfom of
the bath symmetrized autocorrelation function:
$C(t) = 1/2 \, \langle X_{E}(t)X_{E}(0)+ 
X_{E}(0)X_{E}(t)\rangle$.

\section{Single channel use} 
 \label{sec-4}
In this section, we briefly discuss the case in which a single qubit
is sent down the Hamiltonian channel 
(\ref{Hamiltonian}),(\ref{multimode_environment}).
The qubit states before and after the channel transmission read as follows:
 \begin{equation} 
  	\rho^{\cal Q}=\frac{1}{2} \left( 
 	\begin{array}{cc}
 		\rho_{00} \quad & \rho_{01} \\
 		\rho_{10} \quad & \rho_{11} \\
        \end{array}
	\right) \quad \Rightarrow  \quad
        \rho^{\cal Q'}={\cal E}_1(\rho^{\cal Q})= \frac{1}{2} \left( 
 	\begin{array}{cc}
 		\rho_{00}       \quad  & g\rho_{01} \\
 		g\rho_{10} \quad  & \rho_{11} \\
        \end{array}
	\right), 
 \label{dephasing_ch}
 \end{equation}  
 ${\cal E}_1$ is the map for single channel use.
 The dephasing factor $g\in[0,1]$ is deduced 
 from (\ref{general_dephasing_factor}) for $N=1$:
 \begin{equation} 
    g=\exp\Big\{-\lambda^2\,\int_0^\infty \frac{dw}{\pi}
                S(w)\frac{1-\cos(w\tau_p)}{w^2}\Big\}. 
     \label{one_use_dephasing_factor}
 \end{equation}
 It is possible to give a simple representation of the channel action
 in terms of Kraus operators 
 \cite{kn:nielsen-chuang,kn:benenti-casati-strini}:
 \begin{equation}
     {\cal E}_1\big(\rho^{\cal Q}\big)=
     \sum_{m\in \{0,z\}} p_m B_m\, \rho^{\cal Q} \,B_m^\dag, 
     \qquad  B_m=\,\sigma_m
  \label{DephCh_Kraus}
 \end{equation} 
 where $\sigma_0=\openone$, $p_0=(1+g)/2$, 
 and $p_z=(1-g)/2$.

 Using the Kraus form (\ref{DephCh_Kraus}), we can compute  
 $F_e$ for a generic input state \cite{Schumacher96}:
 \begin{eqnarray}
        F_e\big(\rho^{\cal Q}, {\cal E}_1\big)
        &=&\sum_m \big| \mathrm{Tr}_{\cal Q}\big[\rho ^{\cal Q} 
           A_m^{\cal Q}\big]\big|^2 =\frac{1+g}{2}\,+\,\frac{1-g}{2}z^2,
 \label{Fe_OneUse_general}
 \end{eqnarray}
 where the Bloch coordinate $z=\rho_{00}-\rho_{11}$ 
is the expectation value of $\sigma_z$, and $A_m=\sqrt{p_m}B_m$. 
Note that $F_e$ is independent of the coherences $\rho_{01}$.
In particular, we have 
\begin{equation} 
        F_e^{(z=0)}=\frac{1+g}{2}.
 \label{Fe_SingleUse_z0}
 \end{equation}
This case is 
relevant as it takes place when Alice possesses a maximally  
entangled (Bell) pair
${\cal RQ}$ and sends a member of the pair to Bob. The qubit ($\cal Q$) 
sent to Bob is in the maximally mixed state $\rho^{\cal Q}=\frac{1}{2}
\openone$, therefore $z=0$ and Alice and Bob eventually share a pair whose 
fidelity is given by $F_e^{(z=0)}$. This implies that a Bell measurement
able to distinguish the ideally shared Bell state from the other states 
of the Bell basis fails with \textit{error probability} $P_{\mathrm{e}}=1-F_e$.
 
The entropy exchange  $S_e\big(\rho^{\cal Q}, {\cal E}_1\big)$ can be computed 
as the von Neumann entropy of the matrix
 $ W_{mn}= \mathrm{Tr}_{\cal Q}\big[A_m^{\cal Q}\,
   \rho ^{\cal Q}\, A_n^{\dag{\cal Q}}\big]$ \cite {Schumacher96}.
 The eigenvalues of $W$ are  
 $\lambda^{W}_{1,2}=\frac{1}{2}(1\pm \sqrt{g^2+(1-g^2)z^2})$.
 Then the entropy exchange is given by 
\begin{equation}
S_e=S(W)=-\sum_{m=1}^2\lambda^{W}_m\cdot 	
   \log_2\lambda^{W}_m.
\end{equation}
In particular, 
\begin{equation} \label{Se_SingleUes_z0}
S_e^{(z=0)}=H\left(\frac{1+g}{2}\right)=H(F_e^{(z=0)}),
\end{equation}
where $H(x)=-x\log_2 x - (1-x)\log_2(1-x)$ is the Shannon binary entropy.
It is worth noticing that the entropy exchange takes its maximum 
for the completely dephasing channel ($g=0$), for which the entanglement
fidelity $F_e$ is minimum.

 The coherent information 
 is given by
 $      I_c\big(\rho^{\cal Q},{\cal E}_1\big)=
             H(\lambda_1^{\rm out}) - H(\lambda_1^{W}) 
 $,
 where $\lambda_{1,2}^{out}=\frac{1}{2}(1\pm \sqrt{z^2+g^2\gamma^2})$ 
are the eigenvalues of $\rho^{\cal Q'}$. Here $\gamma^2=x^2+y^2$, 
the Bloch coordinates $x=2{\rm Re}(\rho_{01})$ and $y=-2{\rm Im}(\rho_{01})$ 
being the expectation values of $\sigma_x$ and $\sigma_y$.
It is easy to show that the coherent information is maximized by 
the input state $\rho^{\cal Q}=\frac{1}{2}\openone$, that is, 
for $x=y=z=0$. In this case,
\begin{equation} \label{Ic_SingleUes_z0} 
I_c^{(z=\gamma=0)}=1-S_e^{(z=0)}=1-H\left(\frac{1+g}{2}\right).
\end{equation}
It can be proved that (\ref{Ic_SingleUes_z0}) is the quantum capacity
for a memoryless dephasing 
channel \cite{DevetakShor04,Giovannetti05,WolfGarcia07}.

\section{Memory dephasing channel: two channel uses} 
\label{sec-5}
In this section, we consider two channel uses.
Provided  that the time $\tau$ between two channel uses is smaller than the
time scale $\tau_c$ associated with the decay of environmental 
correlation functions, the action of the environment on the second qubit
is related to the action on the first qubit. Therefore, 
${\cal E}_2 \neq {\cal E}_1\otimes {\cal E}_1$, where the superoperator 
${\cal E}_2$ describes the transformation operated by the channel on the
overall two-qubit system.
We show that channel memory can enhance the coherent information.
Moreover, one can exploit memory effects to design suitable 
coding-decoding schemes that improve the faithfulness of  
quantum information transmission.

\subsection{The system transformation}
\label{sec-5.1}
We consider the transmission of two qubits, ${\cal Q}_1$ and ${\cal Q}_2$,
initially in the state $\rho^{{\cal Q}_1{\cal Q}_2}$, with matrix
elements $\rho_{mn}$, $m,n=0,...,3$. The final 
state of the system is
\begin{eqnarray}
  && \rho^{{\cal Q}'_1{\cal Q}'_2} = {\cal E}_2(\rho^{{\cal Q}_1{\cal Q}_2}) 
\,=\,\left( 
     \begin{array}{cccc}
              \rho_{00} \quad &    g\cdot \rho_{01} \quad &  
                                   g\cdot \rho_{02} \quad & h^+\cdot \rho_{03}\\
        g\cdot\rho_{10} \quad &           \rho_{11} \quad &  
                                   h^-\cdot\rho_{12}\quad & g\cdot \rho_{13}  \\
        g\cdot\rho_{20} \quad & {h^-}\cdot\rho_{21} \quad & 
                                          \rho_{22} \quad &  g\cdot \rho_{23} \\
   {h^+}\cdot \rho_{30} \quad &    g\cdot \rho_{31} \quad & 
                                   g\cdot \rho_{32} \quad &     \rho_{33}     \\
        \end{array}
	\right),   
\label{rho_out_two}            
\end{eqnarray}
where the factors $g$ and $h^\pm$  describe the channel effects and the 
noiseless limit is recovered for $g=h^\pm=1$. The last two terms 
are defined as
\begin{eqnarray}
	h^\pm&=&\exp\Big\{-2\lambda^2\,\int_0^\infty 
		\frac{dw}{\pi}S(w)\frac{1-\cos(w\tau_p)}
	{w^2}(1\pm\cos w\tau)\Big\}
        \label{h+-} 
\end{eqnarray}
and are derived from (\ref{general_dephasing_factor}) for
$j_1=j_2=j$, $l_1=l_2\neq j$ ($h^+$) and for
$j_1=l_2=j$, $j_2=l_1\neq j$ ($h^-$). 

In the absence of any memory effects, that is, when the power spectrum 
$S(\omega)$ is white and there is no superposition between the time 
windows when the first or the second qubits are inside the channel
($\tau \geq \tau_p$), we have
$h^\pm=g^2$; therefore, ${\cal E}_2 = {\cal E}_1 \otimes {\cal E}_1$.
In the opposite limiting case of perfect memory, that is, when
$\tau_c\gg \tau,\tau_p$, or alternatively the two time windows of 
the qubit-environment interaction are completely superimposed ($\tau=0$),
we have  $h^+=g^4$ and $h^-=1$. In this limit the subspace  
spanned by the basis $\{\vert 00\rangle\,, \vert11\rangle\}$ 
undergoes a stronger decoherence (with respect to the memoryless case), 
while the subspace spanned by 
$\{\vert 01\rangle\,, \vert10\rangle\}$ is \textit{decoherence free}.

It is convenient to measure the memory between two channel uses
by introducing the memory coefficient $\gamma$ defined as follows:
\begin{equation}
  \gamma=                 
  \int_0^\infty \frac{dw}{\pi} S(w)\frac{1-\cos(w\tau_p)}{w^2}\cos w\tau\,
   \Big/\int_0^\infty \frac{dw'}{\pi}S(w')\frac{1-\cos(w'\tau_p)}{w'^2}.
   \label{gamma}	    
\end{equation}
For a given power spectrum $S(\omega)$ and crossing time $\tau_p$, 
$\gamma$ only depends on the time interval $\tau$, 
and ranges in the interval $[0,1]$. In particular, 
$\gamma=0$ for a memoryless channel 
(as it can be checked by letting $\tau \to \infty$ in the (\ref{gamma})),
while $\gamma=1$ for perfect memory ($\tau=0$ in (\ref{gamma})). 
We can express the dephasing factors $h^{\pm}$ by means of 
the corresponding memoryless value $g^2$ and the memory factor 
$\gamma$:
\begin{equation}
   h^\pm=g^{2(1\pm \gamma)}. 
  \label{h+-gamma}
\end{equation}

\subsection{Kraus representation}
\label{sec-5.2}
In presence of memory ($\gamma>0$), 
the Kraus representation in our Hamiltonian model cannot be trivially derived 
from (\ref{DephCh_Kraus}) by simply concatenating the Kraus operators 
for single channel use with a 
suitable probability distribution:
\begin{equation}
   {\cal E}_2 (\rho^{{\cal Q}_1{\cal Q}_2}) \, 
             \neq \, \sum_{m_1 m_2 \in \{0,z\}} p(m_2,m_1) \,
             B_{m_1}\otimes B_{m_2} \, \rho^{{\cal Q}_1{\cal Q}_2} \,
             B_{m_1}^\dag\otimes B_{m_2}^\dag,
\label{kraus_2}
\end{equation}
where the $B_m$ operators are the same as in (\ref{DephCh_Kraus}).
It is worth noting that in (\ref{kraus_2}) memory could be taken
into account by means of the joint probability $p(m_2,m_1)=p(m_2|m_1)p(m_1)$,
where the conditional probability $p(m_2|m_1)$ can introduce a correlation
between the occurrence of the second qubit operator $B_{m_2}$ and the action
of first qubit operator $B_{m_1}$. 
For simplicity we rename all possible combinations of the $B_m$ operators 
in (\ref{kraus_2}): 
  $K_0 \equiv B_0 \otimes B_0$,
  $K_1 \equiv B_0 \otimes B_z$,
  $K_2 \equiv B_z \otimes B_0$ and 
  $K_3 \equiv B_z \otimes B_z$.
It turns out that the Kraus representation for the map eq. (\ref{rho_out_two})
 requires two other operators
 $K_4 \equiv \frac{1}{2}(K_1+K_2)$ and 
 $K_5 \equiv \frac{1}{2}(K_0-K_3)$:
\begin{equation}
   {\cal E}_2 (\rho^{{\cal Q}_1{\cal Q}_2}) \, = \, \sum_{m=0}^5 p_{K_{m}} \,
             K_{m} \, \rho^{{\cal Q}_1{\cal Q}_2} \,
             K_{m}^\dag\
\label{kraus-2-memory}
\end{equation}
where 
$p_{K_{0}}=\frac{1}{4}(1+2g+h^+)$,
$p_{K_{1}}=p_{K_{2}}=\frac{1}{4}(1-h^-)$, 
$p_{K_{3}}=\frac{1}{4}(1-2g+h^+)$ and
$p_{K_{4}}=p_{K_{5}}=\frac{1}{4}(h^--h^+)$.
For $\gamma=0$, (\ref{kraus-2-memory}) is exactly equivalent to
concatenating twice (\ref{DephCh_Kraus}), namely 
${\cal E}_2={\cal E}_1\otimes {\cal E}_1$. 
The behaviour of the (\ref{kraus_2}) and (\ref{kraus-2-memory}) are very different.
In fact for perfect memory the first Kraus representation,
in which we have to set $p(m_2|m_1)=\delta_{m_2,m_1}$, 
generates two decoherence free subspaces:  
$\{\vert 00\rangle\,, \vert11\rangle\}$ as well as
$\{\vert 01\rangle\,, \vert10\rangle\}$.

\subsection{Entanglement fidelity, entropy exchange and coherent information}
\label{sec-5.3}
The purpose of this section is to show that memory effects can improve the 
channel performance. We consider diagonal input states of the type 
\begin{equation} 
   \rho^{{\cal Q}_1 {\cal Q}_2}= \frac{1}{4}\big[
   p\big(\vert 01\rangle\langle01\vert 
            + \vert 10\rangle\langle10\vert \big) +
     q\big(\vert 00\rangle\langle00\vert 
            + \vert 11\rangle\langle11\vert \big)
     \big].
 \label{rho_pq}
\end{equation}
This density operator describes a mixture in which we have, with probability
$p$, the unpolarized state of the
subspace spanned by $\{\vert 01\rangle\,, \vert10\rangle\}$  and,
with probability $q=1-p$, the unpolarized state
of the subspace spanned by $\{\vert 00\rangle\,, \vert 11\rangle\}$.
One can tune $p$ to increase the weight of the first subspace which -
in the presence of memory - is protected against noise.
On the other hand, the amount of input information, measured by the von
Neumann entropy $S\big( \rho^{{\cal Q}_1 {\cal Q}_2} \big)$, is maximal
when $p=q$. 
We can purify system 
${\cal Q}_1 {\cal Q}_2$ by means of a two-qubit 
reference system ${\cal R}_1 {\cal R}_2$.
We choose, for the system ${\cal R}_1 {\cal R}_2 {\cal Q}_1 {\cal Q}_2$, 
the following pure state:
\begin{eqnarray}
 |\psi^{ {\cal R}_1 {\cal R}_2 {\cal Q}_1 {\cal Q}_2 } \rangle =
       \sum_{i,j=0}^{1} c_{ij}\, | ij^{{\cal R}_1 {\cal R}_2}\rangle 
			         | ij^{{\cal Q}_1 {\cal Q}_2}\rangle, 
			  \qquad
	c_{ij}=
        \left\{ 
          \begin{array}{cc}
      	    \sqrt{p/2} \quad \textrm{if} \quad ij = {01, 10},   \\ 
            \sqrt{q/2} \quad \textrm{if} \quad ij = {00, 11}.   \\ 
          \end{array}
        \right.        
        \label{rho_pq_purification}
\end{eqnarray}
After the qubits ${\cal Q}_1$ and ${\cal Q}_2$ have crossed  
the channel,
the system ${\cal R}_1 {\cal R}_2 {\cal Q}_1 {\cal Q}_2$ 
is described by the density operator
 \begin{equation}
\begin{array}{c}
\rho^{{\cal R}_1 {\cal R}_2 {\cal Q}_1' {\cal Q}_2'}=
      \big({\cal I}^{{\cal R}_1{\cal R}_2}\otimes{\cal E}_2^{{\cal Q}_1{\cal Q}_2}\big)
\big(\vert  \psi^{{\cal R}_1{\cal R}_2{\cal Q}_1{\cal Q}_2}\rangle
            \langle\psi^{{\cal R}_1{\cal R}_2{\cal Q}_1{\cal Q}_2}\vert \big)
\nonumber\\
\nonumber\\
      =\frac{1}{4}
        \left( 
          \begin{array}{ccccccc}
            q      & \ldots & g\sqrt{pq} & \ldots & g\sqrt{pq} & \ldots &  qh^+ \\
            \vdots & \vdots & \vdots & \vdots & \vdots & \vdots &  \vdots  \\
	    g\sqrt{pq} & \ldots & p & \ldots & ph^- & \ldots & g\sqrt{pq}  \\
            \vdots & \vdots & \vdots & \vdots & \vdots & \vdots &  \vdots  \\
            g\sqrt{pq} & \ldots & ph^- & \ldots & p & \ldots & g\sqrt{pq}  \\
            \vdots & \vdots & \vdots & \vdots & \vdots & \vdots &  \vdots  \\
            qh^+ & \ldots & g\sqrt{pq} & \ldots & g\sqrt{pq} & \ldots &  q \\
          \end{array}
        \right),
\end{array}
 \label{rho_pq_out}
\end{equation}
where the dots stand for zeros (the matrix has dimension $2^4\times 2^4$).

\begin{figure}
       \begin{center}  
       \resizebox{0.95\columnwidth}{!}{\includegraphics{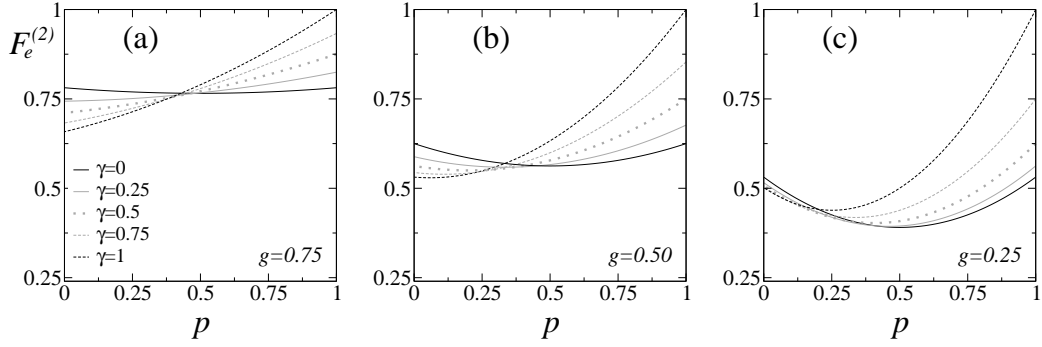}}
       \caption{Entanglement fidelity for two channel uses and
                input state (\ref{rho_pq}), for different values      
                of the dephasing factor $g$ and the memory
                factor $\gamma$.} 
       \label{Fe2pq_fig}
       \end{center}  
\end{figure}

The entanglement fidelity is 
\begin{equation}
 	F_e^{(2)}\equiv F_e\big(\rho^{{\cal Q}_1 {\cal Q}_2},
             {\cal E}_2\big)\,
        =\, 
        \langle \psi^{{\cal R}_1 {\cal R}_2 {\cal Q}_1 {\cal Q}_2}|
         \rho^{{\cal R}_1 {\cal R}_2 {\cal Q}_1' {\cal Q}_2'}
        |\psi^{{\cal R}_1 {\cal R}_2 {\cal Q}_1 {\cal Q}_2}\rangle
\end{equation}
and after some calculations we obtain
\begin{equation}
 F_e^{(2)}=\frac{1}{2}\big[q^2(1+h^+)+p^2(1+h^-)+4gpq\big].
 \label{Fe_2_pq}
\end{equation}
In Fig~\ref{Fe2pq_fig}, we show several plots of the 
two-qubit entanglement fidelity (\ref{Fe_2_pq}). 
We can say that memory effects
always improve this quantity, provided that the factor $p$
is properly chosen. This effect
is more evident for strong dephasing (see Fig.~\ref{Fe2pq_fig}(c)).

\begin{figure}
       \begin{center}  
       \resizebox{0.95\columnwidth}{!}{\includegraphics{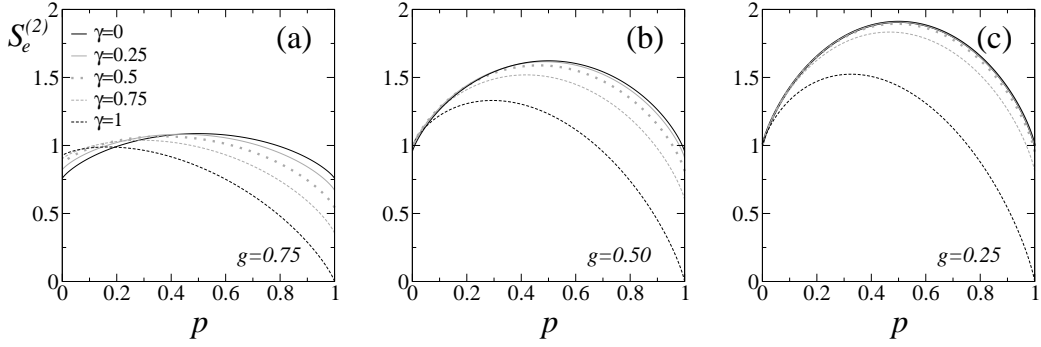}}
       \caption{Same as in Fig.~\ref{Fe2pq_fig} but for the entropy exchange.}
       \label{Se2pq-fig}
       \end{center}
\end{figure}

We now turn our attention to the entropy exchange
\begin{equation}
 	S_e^{(2)}\equiv S_e\big(\rho^{{\cal Q}_1 {\cal Q}_2},
		{\cal E}_2\big)\,
        =\, S\Big(\rho^{{\cal R}_1 {\cal R}_2 
			{\cal Q}_1' {\cal Q}_2'}\Big) 
        =\,
  	-\sum_m \lambda_m\, \log_2 \lambda_m,
  \label{Se_2_pq}
\end{equation}
where the non-trivially equal to zero eigenvalues $\lambda_m$ of 
the output density operator (\ref{rho_pq_out}) are given by 
\begin{eqnarray}
 \lambda_{1}  \,=\, \frac{1}{2}q(1-h^+),\qquad
 \lambda_{2}  \,=\, \frac{1}{2}p(1-h^-)\qquad \textrm{{and}} \qquad
 \lambda_{3,4}  \,=\, \frac{1}{4}\big(1+h^++h^- \pm \Delta_{34} \big), 
 \label{2Qrho_out_Eigenvalues}
\end{eqnarray}
with
\begin{eqnarray}
\Delta_{34} & = & \sqrt{(1+qh^+ + ph^-)^2+4pq[4g^2-(1+h^+)(1+h^-)]}.
\end{eqnarray}
As shown in Fig.~\ref{Se2pq-fig},
memory effects lower the entropy exchange,
that is, memory reduces to some extent the information on the system 
acquired by the channel-environment.
These plots are complementary to those in Fig.~\ref{Fe2pq_fig}:
as expected from the quantum Fano inequality \cite{Schumacher96}, the 
information exchanged with the environment is small when
the disturbance of the state is small, namely $S_e$ is close 
to zero when $F_e$ is close to one. 
The numerical data of Figs.~\ref{Fe2pq_fig}-\ref{Se2pq-fig} 
also show that $S_e$ is close 
to one when $F_e$ is close to zero.

\begin{figure}
       \begin{center}  
       \resizebox{0.95\columnwidth}{!}{\includegraphics{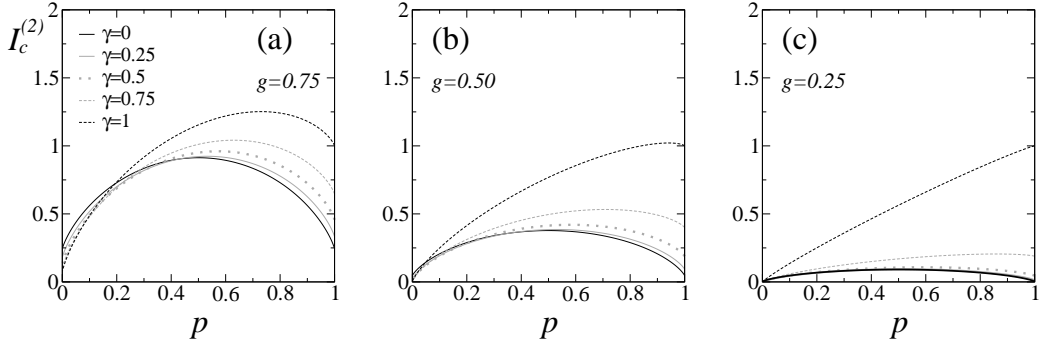}}
       \caption{Same as in Fig.~\ref{Fe2pq_fig} but for the 
       coherent information.}
       \label{Ie2pq-fig}
       \end{center}
\end{figure}

\begin{figure}
       \resizebox{0.95\columnwidth}{!}{\includegraphics{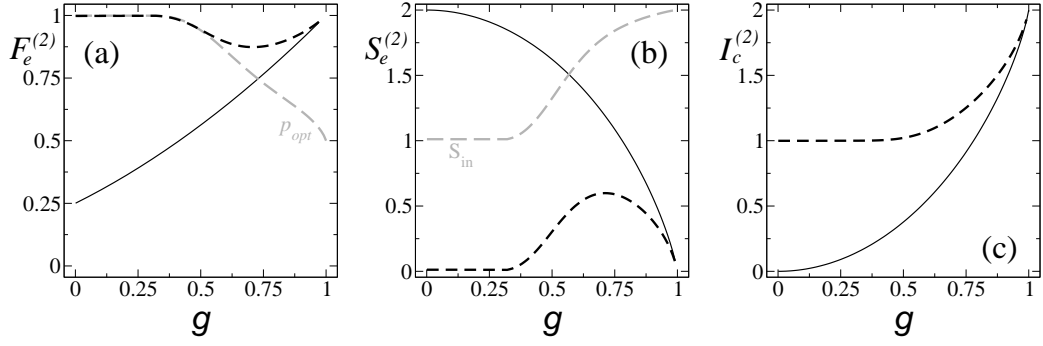}}
       \caption{Plots of (a) entanglement fidelity, (b) entropy exchange
                and (c) coherent information 
                as functions of $g$ for the input state (\ref{rho_pq}),
                where $p$ is chosen to maximize the
                coherent information (black dashed curves); we 
                plot the corresponding curves for the memoryless case 
                (black full curves).
                We also show in panel (a) $p_{\rm opt}$ as a 
                function of $g$ and in panel (b)
                the corresponding value 
                of the input state entropy $S_{\rm in}$ (dotted grey curves).
 		} 
      \label{Opt}
\end{figure}

Finally we turn to the coherent information
\begin{eqnarray}
	I_c^{(2)}\equiv I_c\big(\rho^{{\cal Q}_1 {\cal Q}_2},
		{\cal E}_2\big) 
        \, &=& \,
 	S\big(\rho^{{\cal Q}_1' {\cal Q}_2'}\big) \, - \,
        S_e^{(2)}.
 \label{Ie_2}
 \end{eqnarray} 
Since the coherence terms in $\rho^{{\cal Q}_1 {\cal Q}_2}$
are equal to zero, the dephasing channel does not change this state
and therefore
$S(\rho^{{\cal Q}_1' {\cal Q}_2'})=
S(\rho^{{\cal Q}_1 {\cal Q}_2})=
-p\log_2\frac{p}{2}-q\log_2\frac{q}{2}$ and
\begin{eqnarray} 
	I_c^{(2)}=-p\log_2\frac{p}{2}-q\log_2\frac{q}{2} - \,S_e^{(2)}.
 \label{Ie_2_pq}
\end{eqnarray} 
We can see from fig. \ref{Ie2pq-fig} that coherent information is 
highly sensitive
to memory effects. This sensitivity depends on the dephasing factor $g$.
When dephasing is strong ($g$ close to zero), strong memory effects are 
required to enhance the coherent information (see Fig.~\ref{Ie2pq-fig}(c)),
while for weaker dephasing (panels (a) and (b)
in the same figure)
it is possible to tune $p$ to obtain noticeable enhancements also with
weak memory effects.
It is interesting to examine the limiting case of perfect memory ($\gamma=1$)
and to find, as a function of $g$, the value $p_{\rm opt}$ that
maximizes the coherent information achievable for input 
state as in (\ref{rho_pq}). Fig~\ref{Opt} shows that also in the regime of
strong dephasing ($g<0.5$) reliable transmission can be obtained,
provided we exploit the decoherence-protected subspace 
$\{\vert 01\rangle\,, \vert10\rangle\}$, that is,
$p_{\rm opt}$ is close to one.

\subsection{Coding-decoding scheme taking advantage of channel memory}
\label{sec-5.4}
\begin{figure} 
       \begin{center}
       \resizebox{0.8\columnwidth}{!}{\includegraphics{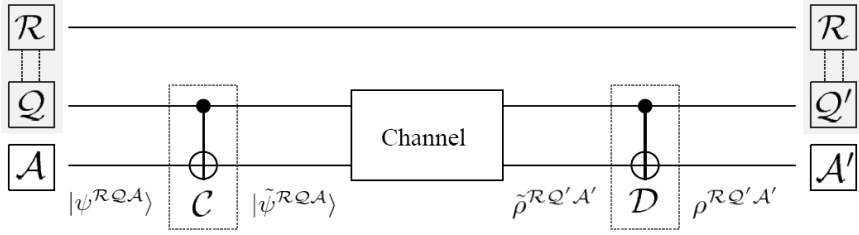}}
       \caption{Sketch of a coding-decoding scheme taking 
                advantage of the correlation between two channel uses.	
 		} 
 \label{coding-fig}	
 \end{center} 		 
 \end{figure}

We show that memory effects can be used to preserve entanglement
in quantum information transmission. 
In particular, we consider an entanglement sharing protocol: Alice
wish to send one qubit of a Bell pair (qubits ${\cal R}$
and ${\cal Q}$) to Bob. The quantum channel 
(\ref{Hamiltonian}),(\ref{multimode_environment})
randomizes the phase between the sent qubit (${\cal Q}$) and
the reference one (${\cal R}$). 
In order to take advantage of memory effects, we follow
the strategy sketched in Fig~\ref{coding-fig}. 
We encode the sent qubit in a two-qubit system
whose state resides in the subspace 
(spanned by $\{\vert 01\rangle\,, \vert10\rangle\}$) resilient
to errors in the presence of memory effects.
 
Without loss of generality we assume
that the initial Bell state of the entangled pair is 
$|\psi^{\cal RQ}\rangle\,=\frac{1}{2}\big(|00\rangle+|11\rangle\big)=
|\phi^+\rangle$. 
The coding-decoding protocol is performed in the following way: 
\newline \textbf{(a)} We prepare an ancillary qubit $\cal A$ in the state 
    $|1\rangle^{\cal A}$, such that the whole system ${\cal RQA}$ is initially
    in the state
    \begin{eqnarray} \label{quec1_a}
      |\psi^{\cal RQA}\rangle &=&|\psi^{\cal RQ}\rangle\otimes|1\rangle^{\cal A}=
      \frac{1}{2}\big(|001\rangle\,+\, |111\rangle \big).
    \end{eqnarray}
\newline \textbf{(b)} The encoding operation $\cal C$ is a controlled-not gate 
     acting on the system $\cal QA$, where $\cal Q$ is the control qubit:
     \begin{eqnarray} \label{quec1_b}
       |\tilde{\psi}^{\cal RQA}\rangle
        =\Big({\openone}^{\cal R}\otimes\textrm{CNOT}^{\cal QA}\big)
 	      \big(|\psi^{\cal RQA}\rangle\Big)
        =\frac{1}{2}\big(|001\rangle\,+\, |110\rangle \big). 
          \end{eqnarray}
    As a result, we encode the system $\cal RQA$ 
    into a GHZ state, in such a way that the subsystem ${\cal Q}{\cal A}$
    resides 
    in the subspace spanned by $\{\vert 01\rangle\,, \vert10\rangle\}$. 
    This is the
    case when the entangled state $|\psi^{\cal R \cal Q}\rangle$ is any 
    state of the Bell basis: 
    \begin{eqnarray}
         |\phi^{\pm}\rangle =\frac{1}{\sqrt{2}}
         \big(|00\rangle\pm|11\rangle\big)\longrightarrow 
            \frac{1}{\sqrt{2}}\big(|001\rangle\, \pm \, |110\rangle\big), 
            \nonumber\\
         |\psi^{\pm}\rangle =\frac{1}{\sqrt{2}}
         \big(|01\rangle\pm|10\rangle\big)\longrightarrow 
             \frac{1}{\sqrt{2}}\big(|010\rangle\, \pm \, |101\rangle\big).
    \end{eqnarray}
\newline \textbf{(c)} We send qubits ${\cal Q}$ and ${\cal A}$ 
          through the channel.
          We call $\tilde{\rho}^{\cal RQ'A'}$ the density operator describing
          the state that arises from the channel transmission:
          \begin{equation}
                \tilde{\rho}^{\cal RQ'A'}=
			\big({\cal I}^{\cal R}\otimes{\cal E}^{\cal QA}\big)
                              \big(|\tilde{\psi}^{\cal RQA}\rangle
                              \langle\tilde{\psi}^{\cal RQA}|\big).
          \end{equation}
\newline \textbf{(d)} The decoding operation $\cal D$ 
         extracts the state of system ${\cal RQ}$ from the one of $\cal RQA$.
         To this aim we apply another controlled-not gate 
         to the system $\cal QA$, where $\cal Q$ is - as above - the 
         control qubit. This operation disentangles systems $\cal RQ$ and 
         $\cal A$:
	 \begin{eqnarray} \label{quec1_dec}           
	          \rho^{\cal RQ'A'} \,=\,
                 \Big({\openone}^{\cal R}\otimes\textrm{CNOT}^{\cal QA}\Big)
                          \, \tilde{\rho}^{\cal RQ'A'} \,
                    \Big({\openone}^{\cal R}\otimes\textrm{CNOT}^{\cal QA}\Big)=
                 \rho^{\cal RQ'} \otimes |1\rangle^{\cal A}, 
          \end{eqnarray}
         where 
         \begin{equation} \label{reducedRHOqec}
          \rho^{\cal RQ'}\,=\, \frac{1}{2}      
          \big(\vert 00\rangle\langle00\vert +\vert 11\rangle\langle11\vert\big) 
           + \frac{h^-}{2}
          \big(\vert 00\rangle\langle11\vert +\vert 11\rangle\langle00\vert\big).
        \end{equation}
\newline The fidelity of the final state 
$\rho^{\cal RQ'}$ is       
\begin{equation}
       F= \langle \phi^{\cal RQ}| \rho^{\cal RQ'} 
                    |\phi^{\cal RQ}\rangle\,=\, \frac{1+h^-}{2}.
\end{equation}
This is also the entanglement fidelity $F_e^c$ when the initial state of
$\cal Q$ is $\rho^{\cal Q}=\frac{1}{2}{\openone}$ and when 
the above coding-encoding scheme is used.
We compare this value with the entanglement fidelity 
(\ref{Fe_SingleUse_z0}), obtained when the qubit ${\cal Q}$ is simply
sent down the channel.
Therefore, the above coding-encoding strategy is useful when memory effects
are strong enough, namely when
\begin{equation}
    F_e^c \geq F_e \quad \Rightarrow
    \quad h^-=g^{2(1-\gamma)}\geq g \qquad \Longrightarrow \quad 
\gamma \geq 0.5.
\end{equation}
In spite of its simplicity, coding/deconding schemes similar to the one
described in this section can be useful to protect information in 
the presence of memory effects \cite{Yamamoto}.

\section{Conclusions}

In this paper we have shown that,
already for two channel uses, memory effects
can deeply modify the behaviour of a quantum Hamiltonian dephasing 
channel with respect to the memoryless limit. 
It is relevant that memory, already with two channel uses, 
can be used to enhance the channel capability to transmit 
coherent quantum information. This result may be of interest for
present-day few-qubit experimental implementations of quantum hardware.

\begin{acknowledgement}
We thank A. Mastellone and E. Paladino for discussions.
G.B. acknowledges support from MIUR-PRIN 2005 (2005025204).
A.D. and G.F. acknowledge support from the EU-EuroSQIP (IST-3-015708-IP) and
MIUR-PRIN2005 (2005022977).
\end{acknowledgement}

\end{document}